# Long-lived excitons in GaN/AlN nanowire heterostructures

M. Beeler,[1,2] C. B. Lim,[1,2] P. Hille,[3] J. Bleuse,[1,2] J. Schörmann,[3] M. de la Mata,[4] J. Arbiol,[4,5] M. Eickhoff,[3] and E. Monroy[1,2]

[1] *Université Grenoble Alpes, 38000 Grenoble, France*

[2] *CEA-Grenoble, INAC, SP2M / NPSC, 17 av. des Martyrs, 38054 Grenoble, France*

[3] *I. Physikalisches Institut, Justus-Liebig-Universität Gießen, Heinrich-Buff-Ring 16, D-35392 Gießen, Germany*

[4] *Institut de Ciencia de Materials de Barcelona, ICMAB-CSIC, Campus UAB, 08193 Barcelona, Catalonia, Spain*

[5] *ICREA and Institut Català de Nanociència i Nanotecnologia (ICN2), 08193 Barcelona, Catalonia, Spain*

**ABSTRACT.** GaN/AlN nanowire heterostructures can display photoluminescence (PL) decay times on the order of microseconds that persist up to room temperature. Doping the GaN nanodisk insertions with Ge can reduce these PL decay times by two orders of magnitude. These phenomena are explained by the three-dimensional electric field distribution within the GaN nanodisks, which has an axial component in the range of a few MV/cm associated to the spontaneous and piezoelectric polarization, and a radial piezoelectric contribution associated to the shear components of the lattice strain. At low dopant concentrations, a large electron-hole separation in both the axial and radial directions is present. The relatively weak radial electric fields, which are about one order of magnitude smaller than the axial fields, are rapidly screened by doping. This bidirectional screening leads to a radial and axial centralization of the hole underneath the electron, and consequently, to large decreases in PL decay times, in addition to luminescence blue shifts.

**Introduction.** Carrier lifetimes are directly related to detection, spontaneous emission, and stimulated emission efficiencies. Short radiative lifetimes in the picosecond or nanosecond range are useful in light emitters to compete with non-radiative recombination processes. On the other hand, long carrier lifetimes (microsecond) increase the collection probability of photogenerated carriers in solar cells or photodetectors, and can enhance the population inversion in lasers.

One approach to tune the band-to-band radiative time is controlling the electron-hole spatial separation. The carrier separation is achieved through the use of type II heterojunctions,[1,2] or through the introduction of internal electric fields via doping or compositional gradients. In the case of polar materials, such as wurtzite III-nitride or II-oxide semiconductors,[3,4] internal electric fields appear spontaneously in heterostructures due to the polarization difference between binary compounds.[5] In particular, adding up spontaneous and piezoelectric polarization, AlN/GaN quantum wells present an internal electric field on the order of 10 MV/cm,[6] which leads to efficient electron-hole separation along the polar <0001> axis, and considerably increases the band-to-band radiative recombination time.[7–13]

Further control of the carrier lifetime in typical device architectures can be achieved by confining carriers in an additional dimension, i.e. using three-dimensional (3D) nanostructures like quantum dots. Excitons trapped in such quantum nanostructures are efficiently isolated from dislocation or surface related non-radiative recombination centers,[14] which attenuates the quenching of the carrier lifetime with temperature. The synthesis of quantum dots as nanodisks inserted in nanowires (NWs) offers an exciting alternative to avoid the inherent constraints of Stranski-Krastanov growth. In NW geometries, the 3D elastic strain relaxation via the surface in the form of plane bending[15] permits a wider range of quantum dot sizes and compositions before forming misfit dislocations, i.e. plastic relaxation.[15,16] In the case of GaN/AlN quantum dots or nanodisks, the large band offsets (~1.8 eV in the conduction band[17]) provide efficient exciton confinement, so that the observed long (microsecond) photoluminescence (PL) decay times[9,12] can persist up to room temperature.[18]

Micro-PL studies of GaN nanodisks in NWs show excitonic emission lines stemming from discrete levels, as verified by photon-correlation measurements,[19] which confirm their quantum dot-like behavior. The PL spectral positions present signatures of both quantum confinement and of the polarization-induced Stark effect.[16,19–23] However, the spectral shift associated to the Stark effect is smaller in nanodisks than in quantum wells,[16,19,23] which has been attributed to dislocations,[16] to the surface band bending,[24] and to the 3D strain configuration.[19,25] Studies of the PL decay times in GaN NWs show exponential or bi-exponential PL decays with sub-nanosecond characteristic times.[26,27] In GaN/AlN NW heterostructures, sub-nanosecond PL decay times have been reported in the case of small nanodisks (~1 nm), where the emission properties are dominated by the carrier confinement and the quantum confined Stark effect is still negligible.[19] In the case of nanodisks larger than 2 nm, where the emission becomes dominated by the carrier separation due to the polarization-induced internal electric field, time-resolved PL reports are so far limited to the descreening of the polarization-induced internal electric field, in the tens of nanoseconds range,[16,28] where the use of excitations rates larger than 50 MHz (less than 20 ns between pulses) has hindered the observation of the characteristic PL decay times.

In this work, we report the observation of long-lived (microsecond) excitons in GaN/AlN NW heterostructures at room temperature, and we present a comprehensive analysis of the carrier dynamics by combining continuous-excitation and time-resolved PL measurements with 3D calculations of the electronic structure. Electric fields in the axial and radial directions translate into in-plane carrier separations that govern the carrier dynamics at low dopant concentrations, and are rapidly screened by Ge doping in the nanodisks. As a result, screening effects in nanodisks are significantly stronger than in planar structures. By varying the Ge concentration in the GaN nanodisks, the lifetime of photo-induced carriers can be varied by nearly two orders of magnitude.

**Experimental.** N-polar AlN/GaN NW heterostructures were synthesized by plasma-assisted molecular-beam epitaxy (PAMBE) on Si(111) substrates at a growth temperature of ~790°C. At this temperature and under N-rich growth conditions, PAMBE is known to produce N-polar catalyst-free GaN NWs with a radius in the range of a few tens of nanometers.[29–34] The structures under study consist of a non-intentionally doped (n.i.d.) GaN NW base with a length of 600 nm and a radius ranging from 25 to 40 nm, followed by 40 periods of GaN:Ge/AlN (nominally 4 nm/ 4 nm) nanodisks, and a 20-nm-thick n.i.d. GaN cap layer. The GaN nanodisks were doped with Ge, using a beam equivalent pressure ranging from 0.5 to $1.5\times10^{-9}$ mbar. Ge was used as a dopant instead of Si as it introduces less strain in GaN being similar in size to Ga,[36,37] and the change of the NW aspect ratio is negligible even for high Ge doping levels ($3.3\times10^{20}$ cm$^3$).[38] The dopant concentrations of the samples under study, estimated from secondary ion mass spectroscopy measurements in reference samples,[38] are summarized in Table I.

Structural and morphological characterization of the heterostructures was performed by high-resolution transmission electron microscopy (HRTEM) and high angle annular dark field (HAADF) scanning transmission electron microscopy (STEM) using a FEI Tecnai F20 field emission gun microscope operated at 200 kV. For microscopy studies, the NWs were directly scratched from the substrate with a holey carbon TEM grid. The periodicity of the samples was analyzed by high resolution x-ray diffraction (HRXRD) using a PANalytical X'Pert PRO MRD system.

PL spectra were obtained by exciting with a continuous-wave frequency-doubled Ar laser ($\lambda = 244$ nm), with an excitation power around 50 µW focused on a spot with a diameter of ~100 µm, giving a power density of about 7 kW/m$^2$. The emission from the sample was collected by a Jobin Yvon HR460 monochromator equipped with an ultraviolet-enhanced charge-coupled device (CCD) camera. In the case of time-resolved PL, samples were excited using a frequency-tripled Ti:sapphire laser ($\lambda = 270$ nm) with pulse width of 200 fs. This laser was augmented with a cavity damper section with a base pulse repetition rate of 54 MHz. This allowed the period between pulses to be varied from 20 ns to 5 µs. The excitation power was about 500 µW. The luminescence was dispersed by a Jobin Yvon Triax320 monochromator and was detected by a Hamamatsu C−5680 streak camera.

**Results.** Figure 1(a) shows an HAADF image of GaN NWs containing an n.i.d. GaN base, followed by the GaN/AlN heterostructure and the GaN cap. The different GaN:Ge and AlN sections can be easily distinguished by the image contrast, which scales with the atomic number of the observed material. A magnified detail of the first GaN:Ge/AlN periods near the base are displayed in temperature color in Figure 1(b). No trace of GaN-AlN interdiffusion is appreciated in the images. Statistics performed on the nanodisks and barrier thicknesses are in good agreement with the nominal values. Figure 1(c) displays a HRTEM image of three GaN:Ge nanodisks embedded in AlN barrier material. In the HRTEM image shown, the darker contrast corresponds to the GaN:Ge insertions while the brighter lattice contrast is the AlN barrier, evidencing the presence of an AlN shell with thickness roughly equal to the size of the barriers. This shell is generated by direct deposition of the impinging Al atoms due to the low Al diffusion length at this growth temperature.[16,39–42] Furthermore, the GaN/AlN interfaces often present {1−102} facets close to the NW sidewalls, highlighted by a dashed line in Figure 1(c), due to the plane bending phenomena related to the elastic strain relaxation.[15,16,25]

TEM images provide a local view of selected NWs, whereas HRXRD measurements give a structural assessment of the NW ensemble. Figure 1(d) depicts the ω–2θ scans of the (0002) x-ray reflections of one of the samples under study. From the satellites of the GaN/AlN superlattice reflection, the superlattice periods are extracted and summarized in Table I. These superlattice periods were in good agreement with those measured locally by means of TEM.

The optical properties of the NW heterostructures were first analyzed by continuous-wave PL spectroscopy. Figure 2(a) shows the low-temperature ($T = 5$ K) emission of the samples, displaying a blue shift with increasing dopant concentration which is attributed to the screening of the internal electric field.[25,28] The peak PL wavelengths are summarized in Table I. Figure 2(b) shows the variation of the normalized integrated PL intensities as a function of temperature. The PL intensities remains almost constant up to about 100 K, after which, at room temperature, they drop to 20 to 40% of their maximum values. This behavior is characteristic of GaN/AlN nanostructures with 3D confinement,[43–46] in contrast to planar structures which generally exhibit a PL quenching of several orders of magnitude at room temperature.[45]

To probe the band-to-band carrier dynamics within this system, the decay of the PL under pulsed excitation was analyzed. As a typical example, Figure 3(a) shows the time-resolved evolution of the PL spectra of sample N3 measured at low temperature ($T = 5$ K).

The emission presents a red shift of 45 nm during the first ~60 ns before a steady-state is obtained, as illustrated in Figure 3(b). This spectral shift is systematic through all the investigated samples and ranges from 0.1 to 0.3 eV, decreasing for increasing doping levels. Following the PL intensity at the maximum of the spectrum as a function of time (trajectory indicated by the red line in Figures 3(a) and (b)), the intensity decay in Figure 3(c) is obtained. Comparing Figures 3(b) and (c), the initial red shift is associated to a pronounced non-exponential drop of the PL intensity during the first ~60 ns, followed by an exponential decay. These PL dynamics are qualitatively the same for all the samples regardless of doping level.

The initial red shift and non-exponential behavior are attributed to the perturbation of the band structure induced by the excitation (screening of the polarization fields), and to band filling, as previously observed in GaN/AlGaN quantum wells.[11] Therefore, in subsequent analyses, only the time constant extracted during the exponential decay regime (dashed line in Figure 3(c)) is addressed. This regime reflects the carrier dynamics of the original band structure, once the photo-induced perturbation from the laser is dissipated.

Figure 4(a) shows the low-temperature ($T = 5$ K) PL evolution for the NW heterostructures with different dopant concentrations. A drastic decrease of the decay time with increasing Ge concentration is observed: the n.i.d. sample displays a decay time on the order of several µs, whereas the decay times for higher dopant concentrations decrease by more than an order of magnitude (to around 100 ns). These decay times are orders of magnitude longer than shown in previous literature reports,[16,28] where the PL decay times were even shorter than those of equivalent quantum well structures. This discrepancy could be explained by the measurement procedure: in refs [16,28], the decay times were estimated from measurements exciting with a pulse repetition rate of 78 MHz (time between pulses =

12.8 ns). Based on the data from ref.[28], the PL from undoped (highest doped) samples would have only dropped to about 60% (14%) of the maximum value before the next laser pulse hit. From their data, a 1/e decay time can be extracted assuming exponential relaxation. However, our experiments prove that the initial relaxation is strongly nonexponential due to the screening of the electric field induced by the laser pulse. With the lower excitation power, after 12.8 ns none of the samples have entered the exponential regime (dashed line in Figure 3(c)), and only the most heavily doped sample would have recovered from the initial blueshift induced by the laser. We therefore conclude that the measurements in ref.[28] provide information mostly about the recovery of the screening of the internal fields induced by the measuring laser. This is in accordance to the decay time's (and the spectral shift's) dependency on laser power reported by Hille et al.[28]

The evolution of the characteristic PL decay time has been analyzed as a function of temperature with the results plotted in Figure 4(b), where the relaxation times were extracted from the exponential decay profiles as indicated in Figure 3(c). In all cases, the PL decay times remain constant (±10%) from 10 K to 300 K, as previously observed in the case of Stransky-Krastanov GaN/AlN quantum dots.[18] This demonstrates that the 3D confinement in the nanodisks efficiently suppresses thermally-activated non-radiative recombination channels up to room temperature. In contrast, temperature-dependent time-resolved measurements of GaN/AlN quantum well samples show a decrease in relaxation time over this temperature range by several orders of magnitude, as described in ref [18] and illustrated in Figure 4(b).

The thermally stable PL decay time in Figure 4(b) lead us to attribute the thermal quenching of the integrated PL intensity in Figure 2(b) to carrier losses during the relaxation process of the hot photoexcited carriers to the exciton ground states. A simplified view of the process can be provided by the three-level model schematically described in the inset of Figure 2(b). In steady-state conditions the carrier generation rate, $\Phi$, equals the relaxation rate to the exciton emitting state <1> plus the non-radiative recombination rate:

$$\Phi = \frac{n_0}{\tau_{NR}} + \frac{n_0}{\tau_{R0}}, \tag{1}$$

where $n_0$ is the optically excited population of the <0> level, and $\tau_{NR}$ and $\tau_{R0}$ are the characteristic times associated to the non-radiative processes and to the relaxation to the exciton emitting state, respectively. By neglecting non-radiative recombination once the excitons are trapped in the nanodisk, assumption supported by the observation of a PL decay constant with temperature, the PL intensity can be described as:

$$I = \frac{n_1}{\tau_{R1}} = \frac{n_0}{\tau_{R0}} = \frac{\Phi}{1 + \tau_{R0}/\tau_{NR}}, \tag{2}$$

where $n_1$ is the population of the exciton emitting state in the nanodisk and $\tau_{R1}$ is the associated characteristic time.

Assuming that the photogeneration ($\Phi$) is constant with temperature, and that non-radiative processes from state <0> are thermally activated, the PL intensity as a function of temperature, $I(T)$, can be described by:

$$I(T) = \frac{I(T=0)}{1 + a\, exp(-E_a/kT)}, \tag{3}$$

where $E_a$ represents the activation energy of the non-radiative process, $kT$ being the thermal energy, and $a$ being a constant coefficient. Solid lines in Figure 2(b) are fits of the experimental data to eq 3. The extracted values of $E_a$ and $a$ are summarized in Table I. An increase in both parameters with larger doping concentration is observed, which points to an enhanced probability of non-radiative processes with increasing carrier density.

**Discussion.** The strong acceleration of the PL decay with increasing Ge concentration points to a screening of the electric field in the nanodisks that drastically defines the radiative carrier lifetime. The magnitude of this effect is much larger than previously reported for GaN quantum wells.[18,47] Comparing the low-temperature PL lifetimes of n.i.d. quantum wells and nanodisks emitting at approximately the same wavelength, i.e. with the same electron-hole separation in *energy*, the decay time in the case of the nanodisks is significantly longer. This juxtaposition is illustrated in Figure 4(b) for the decay times of quantum wells emitting around 450 nm. This discrepancy points to a significantly larger electron-hole separation in *space*, which implies a different electric field distribution.

In order to understand the electric field distribution leading to this giant screening effect, 3D calculations of the NW strain state, band diagram and quantum confined states were performed using the Nextnano3 software[48] with the material parameters described in ref [49]. The NW was defined as a hexagonal prism consisting of a long (50 nm) GaN stem followed by a sequence of 10 AlN/GaN stacks and capped with 18 nm of GaN. The radius of the GaN base was 20 nm, the growth axis was [000−1] and the sidewall faces were {1−100} planes. The structure was defined on a GaN substrate, to provide a reference in-plane lattice parameter. The GaN stem and the AlN/GaN heterostructure were laterally surrounded by an AlN shell, and the whole structure was embedded in a rectangular prism of air, which permits the elastic deformation of misfit strain. Surface states were modelled as a surface charge density of $2\times10^{12}$ cm$^{-2}$ at the air/semiconductor interfaces.[50] The presence of {1−102} facets in the AlN sections was taken into account, as illustrated in Figure 5(a), which shows a (1−100) cross-section view of 3 nanodisks in the stack.

The 3D strain distribution was calculated by minimizing the elastic energy and applying zero-stress boundary conditions at the surface. The effect of doping on the strain distribution was neglected.[36,37] Figures 5(b) and (c) display (1−100) cross-sectional views of the strain components along the <11−20> direction, $\varepsilon_{xx}$, and <0001> direction, $\varepsilon_{zz}$, for 3 nanodisks in the stack. Regarding the $\varepsilon_{xx}$ component, the center of the disk is compressed by the AlN sections ($\varepsilon_{xx} = -1.29\%$) and there is an elastic relaxation close to the sidewalls. In contrast, the $\varepsilon_{zz}$ strain component is almost zero ($\varepsilon_{zz} = -0.025\%$) along the center of the nanodisk, however near the sidewalls the GaN gets significantly compressed due to the presence of the AlN shell (up to $\varepsilon_{zz} = -2.2\%$). The radial inhomogeneous strain results in non-zero $\varepsilon_{xz}$ and $\varepsilon_{yz}$ shear strain components, as illustrated in Figure 5(d), which in turn leads to radial piezoelectric polarization associated to the non-zero $e_{15}$ piezoelectric constant in the wurtzite lattice. On the other hand, this particular strain distribution results also in an increase of the GaN band gap by ~120 meV when moving from the center of the nanodisk to the sidewalls (data obtained using the deformation potentials from ref [51]).

The strain calculation provides a 3D map of the polarization in the heterostructure. With this input, the nonlinear Poisson equation was solved classically to obtain the 3D band structure of the complete wire. After the Poisson equation was solved in equilibrium, the eigenfunctions were calculated by solving the Schrödinger equation in a quantum region that covered one nanodisk in the center of the NW, including the AlN barriers on the top and bottom. Figures 6(a) and (b) show the conduction and valence band profiles along the [000−1] growth axis along the center of the NW for (a) undoped nanodisks and (b) nanodisks with an *n*-type dopant concentration $N_D = 1.7\times10^{20}\,\text{cm}^{-3}$ (sample N3). In both cases, the polarization-induced internal electric fields result in a sawtooth profile with the electron level shifted towards the bottom of the nanodisk and the hole level towards the top of the nanodisk. In the doped structure, the internal electric field is reduced from 5.9 MV/cm to 2.5 MV/cm due to carrier screening.

Figures 6(c) and (d) show the radial conduction and valence band profiles along the [11−20] axis for undoped and doped ($N_D = 1.7 \times 10^{20}$ cm$^{-3}$) nanodisks. In both cases, the conduction (valence) band profile was taken at the bottom (top) interface of the nanodisk. The squared wavefunctions of the first electron and hole levels are also represented. In the case of undoped nanodisks, the band bending induced by the AlN shell pushes the electrons towards the center of the NW, whereas the radial valence band profile has local maxima near the NW sidewalls. This result is in agreement with calculations by Rigutti *et al.*[52] (GaN/AlGaN NW heterostructures with AlGaN shell) and Rivera *et al.*[53] (GaN/AlGaN NW heterostructures without shell), and in the same line that the calculations of Marquardt *et al.*[54] for InGaN/GaN NW heterostructures. Therefore, in addition to the polarization-induced vertical separation of electron and holes, the 3D geometry of the nanodisks leads to a radial separation of carriers, which explains the delay of the radiative recombination with respect to the quantum well case. Note that the strain-induced enlargement of the GaN band gap at the ~5 nm closest to the nanodisk sidewalls contributes to separate the carriers, particularly the holes, from the core/shell interface. The presence of {1−102} facets at the top interface of the AlN sections, which enlarges the GaN disks close to the surface, does not have a relevant effect on the radial location of the hole. This is because the spontaneous and piezoelectric polarization shifts the hole towards the top of the GaN nanodisk. However, these facets modify the electron wavefunction, which results in a shift of the band-to-band transition by ~300 meV.

At low dopant concentrations (below $10^{19}$ cm$^3$), the electric field along the [11−20] axis at the top GaN/AlN interface, depicted in Figure 6(e), presents a maximum value of ~0.6 MV/cm, i.e. one order of magnitude smaller than the field along [000−1]. Increasing the doping concentration leads to the screening of the lateral electric field, causing the spatial broadening of the electron wave function described in Figure 6(d). In the valence band, the flattening of the potential profile shifts the hole wavefunction towards the center of the NW radially aligning them with the electron wavefunction. The improved electron-hole wavefunction overlap explains the drastic decrease of the radiative recombination lifetime. Figure 7(b) also shows the attenuation of the electric field in the radial direction as a function of the doping concentration. The transition of the hole towards the center of the NW, i.e. the inversion of the electric field sign, takes place for a doping concentration around $3.5 \times 10^{19}$ cm$^{-3}$.

These calculations consider the presence of negatively charged surface states with a density of $2 \times 10^{12}$ cm$^{-2}$. The negatively charged surface attracts the holes, but is not a critical factor to determine the carrier distribution, since the electric field is mostly associated to the piezoelectric phenomena. Figure 7(a) describes the radial electric field in a similar structure without surface charges. A complete suppression of the surface charges lowers the required doping concentration to invert the electric field to $2.0 \times 10^{19}$ cm$^{-3}$ and results in a radial shift of the zero-field position (which corresponds to the maximum of the hole wavefunction) by about 3 nm towards the center of the NW for low doping levels.

The above-described calculations demonstrate that the radial misalignment of the electron and hole wavefunctions is a determining factor for the band-to-band dynamics in GaN/AlN nanodisks. At low dopant levels, the radial electron-hole separation leads to radiative lifetimes that are significantly longer than in GaN/AlN quantum wells emitting at the same wavelength. Upon increasing the dopant concentration, carrier screening leads to a radial centralization of the hole underneath the electron, and a large decrease in the radiative lifetime. These results are also in agreement with the radiation model shown in Figure 2(b), which concurs that with higher centralization of the hole and electron within the nanowire, there will be a higher energetic barrier ($E_a$) for carriers to recombine non-radiatively at the edges of the nanowire.

**Conclusions.** The carrier dynamics in 40-period GaN/AlN (4 nm/4 nm) NW heterostructures have been explored as a function of the Ge dopant concentration in the GaN disks. Long PL decay times, on the order of microseconds, are measured in non-intentionally doped disks and persist up to room temperature. This confirms the efficiency of quantum confinement in the nanodisks to inhibit non-radiative recombination. The long relaxation times are explained as the result of internal electric fields present in the nanodisks, with an axial component in the range of a few MV/cm associated to spontaneous and piezoelectric polarization, and a radial component associated to the radial variation of lattice strain. Simulations show that at low dopant concentrations, a large electron-hole separation in both the axial and radial directions is present, with holes located axially on top of the nanodisk and radially close to the surface, and electrons located axially at the bottom of the nanodisk and radially centered. The relatively weak radial electric fields, calculated to be one order of magnitude smaller than the axial fields, are rapidly screened by doping, which leads to both a radial and axial centralization of the hole underneath the electron. This bidirectional dopant-induced giant screening leads to large decreases in radiative lifetime by about two orders of magnitude, in addition to the luminescence blue shift.

**ACKNOWLEDGEMENTS.** This work is supported by the EU ERC-StG "TeraGaN" (#278428) project, the LOEWE program of excellence of the Federal State of Hessen (project initiative STORE-E), and by the Spanish MINECO MAT2014-51480-ERC (e-ATOM) and Generalitat de Catalunya 2014SGR1638. JA thanks ICN2 Severo Ochoa Excellence Program. MdlM thanks CSIC Jae-Predoc program.

**TABLES**

**Table I:** Characteristics of the GaN/AlN NW heterostructures under study: Germanium beam equivalent pressure (BEP$_{Ge}$) during the nanodisk growth, Ge concentration deduced from reference Ge-doped GaN NW samples measured by time-of-flight secondary ion mass spectroscopy, GaN/AlN period extracted from HRXRD measurements, low-temperature ($T$ = 5 K) PL peak wavelength, and values of the $E_a$ and $a$ parameters in eq 3 extracted from the fits in Figure 2(b).

| Sample | BEP$_{Ge}$ (mbar) | [Ge] (cm$^{-3}$) | Period from HRXRD (nm) | PL peak wavelength (nm) | $E_a$ (meV) | $a$ |
|---|---|---|---|---|---|---|
| N1 | 0 | n.i.d. | 7.5±0.2 | 454 | 40±10 | 9±2 |
| N2 | 5.0×10$^{-10}$ | 9.0×10$^{19}$ | 7.6±0.2 | 429 | 54±10 | 31±5 |
| N3 | 1.0×10$^{-9}$ | 1.7×10$^{20}$ | 7.4±0.2 | 392 | 53±10 | 42±11 |
| N4 | 1.5×10$^{-9}$ | 3.1×10$^{20}$ | 7.4±0.2 | 384 | 68±10 | 131±50 |

**Figure 1**

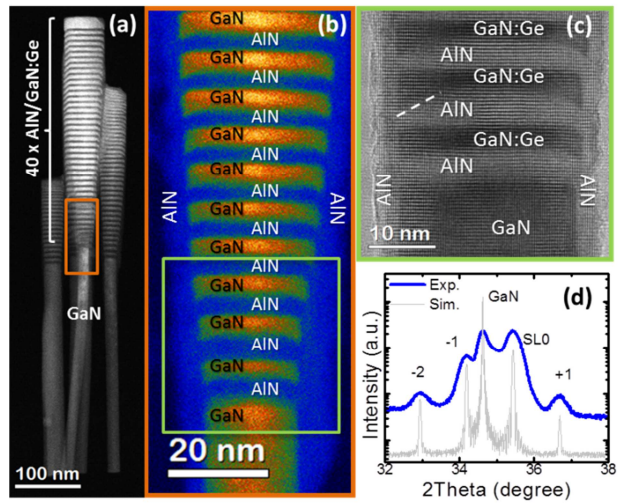

**Figure 1.** (a) HAADF STEM image of the GaN/AlN NW heterostructures. The AlN barriers (darker) and GaN disks (brighter) have nominal thicknesses of 4 nm. (b) Zoom into the squared region in (a) displayed in temperature color code. (c) HRTEM image of the first 3 GaN:Ge disks (near the GaN stem). (d) HRXRD ω−2θ scan around the (0002) reflection of sample N3, together with a simulation. The simulation is down shifted for clarity.

**Figure 2**

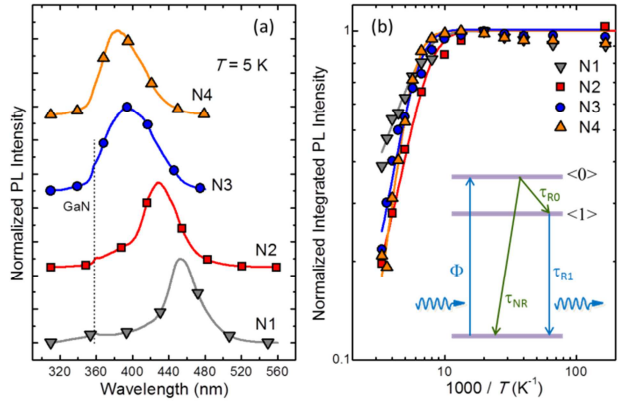

**Figure 2.** (a) Normalized PL spectra of samples N1-N4 measured at low temperature ($T$ = 5 K). The spectra are normalized and vertically shifted for clarity. The dotted vertical line indicates the location of the GaN band gap. (b) Normalized integrated PL intensity of the samples shown in (a) as a function of temperature. Solid lines are fits to eq 3. Inset: Simplified 3-level model of the PL dynamics.

**Figure 3**

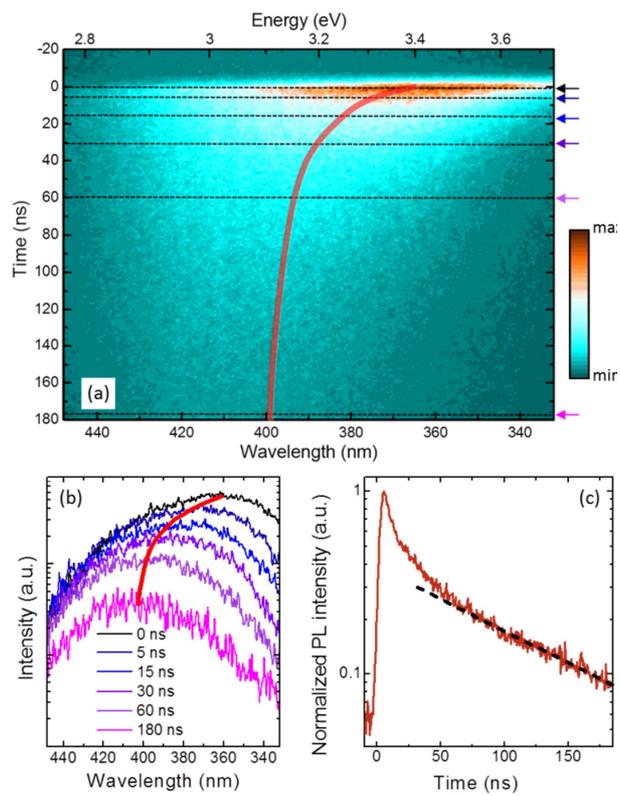

**Figure 3.** (a) Time resolved PL spectra of sample N3. The red stripe follows the intensity maximum as a function of time. (b) Evolution of the PL spectra as a function of time. The time of maximum intensity is taken as $t = 0$. The spectra are acquired with a time integration window of 0.4 ns. (c) Evolution of the PL peak intensity as a function of time. The dashed line is an exponential fit to the PL decay for times longer than 60 ns.

**Figure 4**

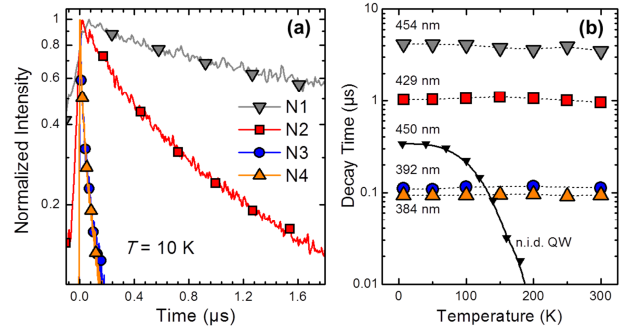

**Figure 4.** (a) PL decay for samples N1-N4 measured at low temperature ($T$ = 10 K). The decays were taken following the method described in Figure 3. (b) PL decay characteristic times extracted from the exponential part of the PL decays (similar to Figure 1(c)), plotted as a function of temperature. The emission wavelengths are indicated in the figure. Superimposed, PL decay times of an n.i.d. GaN/AlN quantum well (labeled n.i.d. QW) emitting at approximately the same wavelength that N1.

**Figure 5**

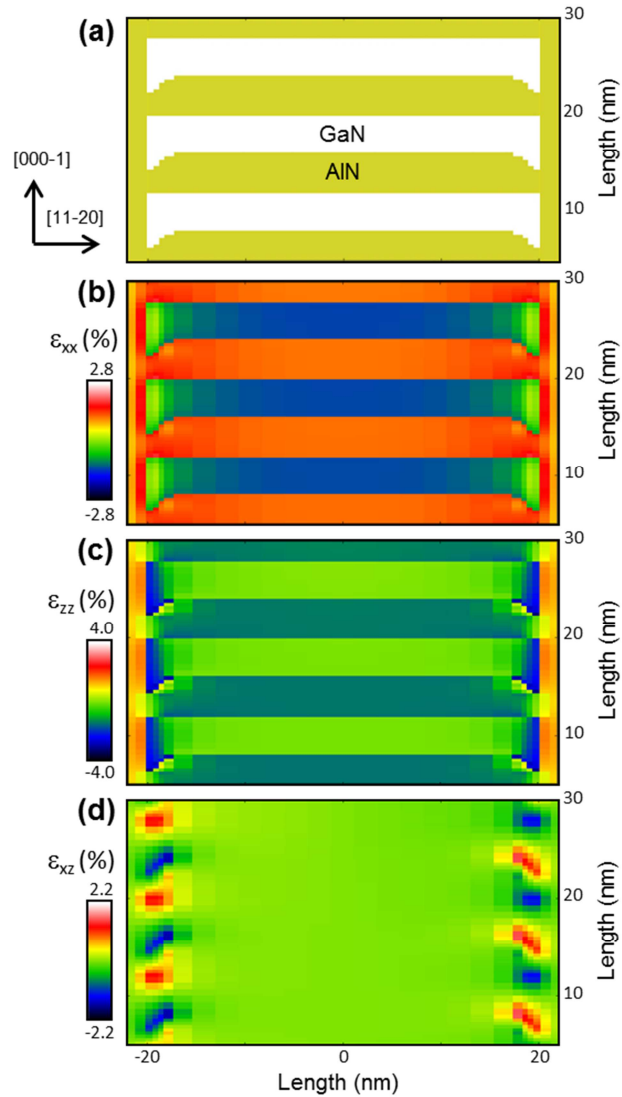

**Figure 5.** (a) Schematic representation of 3 GaN/AlN nanodisks in the center of the nanowire heterostructures, as they are described in the input file for nextnano3. The structure is viewed along the [1−100] plane. White areas correspond to GaN and yellow areas correspond to AlN. (b) Calculation of the $\varepsilon_{xx}$ strain component (strain along [11−20]) for these same disks. (c) Calculation of the $\varepsilon_{zz}$ strain component (along [000-1]). (d) Calculation of the $\varepsilon_{xz}$ shear strain component.

**Figure 6**

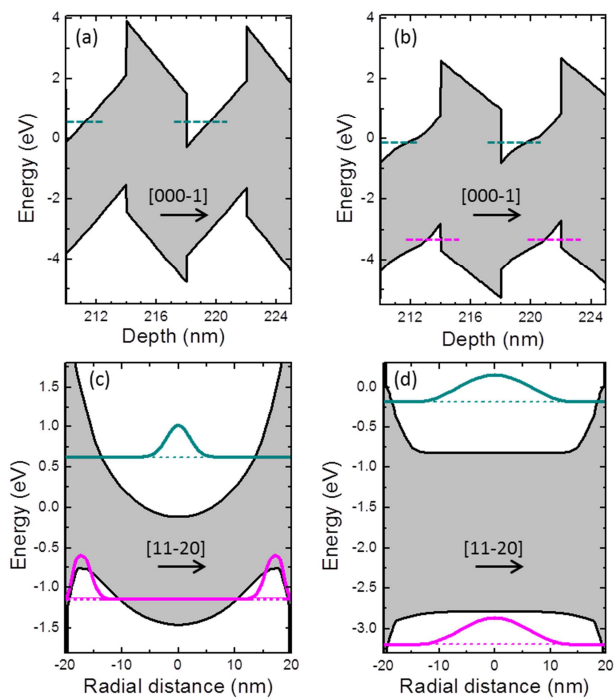

**Figure 6.** (a-d) Calculation of the conduction and valence band profiles and first electronic levels of electrons and holes. (a, b) Conduction and valence band profiles along [000−1] taken at the center of the NW, for (a) undoped nanodisks and (b) Ge-doped nanodisks (sample N3). The ground electron and hole levels are indicated by dashed lines. In (a), the ground hole level is not indicated because the value of the squared wave function along the center of the NW is zero. (c, d) Radial conduction and valence band profiles for the (c) undoped and (d) Ge-doped nanodisk in the center of the stack. Note that the conduction band was taken at the bottom of the disk, while the valence band was taken at the top of the disk. The squared wavefunctions of the ground electron and hole states are indicated in the figures.

**Figure 7**

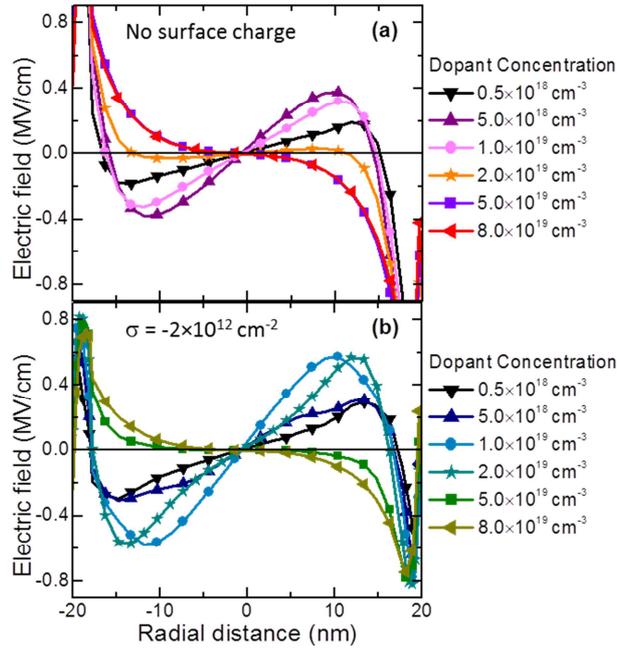

**Figure 7.** Calculation of the electric field along [11−20] at the top of the nanodisk for various doping concentrations in the nanodisks, (a) neglecting surface charges, and (b) with surface charges at a concentration of σ = −2×10$^{12}$ cm$^{-2}$. With higher doping levels, the electric field in the disk is attenuated. Positive electric field implies that it points in the [11−20] direction. The sign of the electric field sees a crossover at a dopant concentration of (a) ~2×10$^{19}$ cm$^{-3}$, and (b) ~4×10$^{19}$ cm$^{-3}$. The magnitude of the radial electric field is (a) 0.42 MV/cm, and (b) 0.58 MV/cm. A null electric field was seen (a) 14 nm and (b) 17 nm from the centre of the wire.